\documentclass[12pt]{article}

\usepackage{graphicx,epsfig,amsmath,amssymb,cite,xspace}
\usepackage{float}

\renewenvironment{thebibliography}[1]
        {\frenchspacing
         \baselineskip=16pt
         \begin{list}{\arabic{enumi}.}
        {\usecounter{enumi}\setlength{\parsep}{0pt}
         \setlength{\leftmargin 12.7pt}{\rightmargin 0pt} %FOR 1--9 ITEMS
         \setlength{\itemsep}{0pt} \settowidth
        {\labelwidth}{#1.}\sloppy}}{\end{list}}

\newcommand{\Jpsi} {\mbox{J\kern-0.05em /\kern-0.05em$\psi$}}
\newcommand{\psip} {\mbox{$\psi'$}}
\newcommand{\Ups}  {\mbox{$\Upsilon$}}
\newcommand{\Upsp} {\mbox{$\Upsilon'$}}
\newcommand{\Upspp}{\mbox{$\Upsilon''$}}

\newcommand{\pt}{\ensuremath{p_{\mathrm{t}}}}

\newcommand{\mrm}{\mathrm}
\newcommand{\dd}{\mrm{d}}
\newcommand{\eg}{{e.g.~\@\xspace}}

\begin{document}

\begin{titlepage}

\begin{flushright}
LIP/96--02\\
11 June 1996
\end{flushright}

\vspace*{1.5cm}
\begin{center}
\large\bf 
HEAVY ION COLLISIONS AT THE LHC\,: \\[0.2cm]
THE ALICE EXPERIMENT
\end{center}
\vspace*{1.2cm}

\begin{center}
\large 
Carlos Louren\c co\\[0.2cm]
CERN, Geneva, Switzerland\\[0.3cm]
(for the ALICE Collaboration)
\end{center}

\vspace*{6cm}
\begin{center}{\bf Abstract}\end{center}

\begin{center}
\begin{minipage}[ht]{0.8\linewidth}
  
  ALICE (A Large Ion Collider Experiment) is a detector designed to
  exploit the physics potential of nucleus--nucleus interactions at
  the LHC. Being a general purpose experiment, it will allow a
  comprehensive study of hadrons, electrons, muons and photons,
  produced in the collision of heavy nuclei, up to the highest
  particle multiplicities anticipated ($\dd N_{\rm ch}/\dd y=8000$).
  In addition to heavy systems (Pb--Pb), we will study collisions at
  smaller energy densities by using lower-mass ions (e.g.
  A$\,\sim40$).  Reference data will be obtained from pp and
  p--nucleus collisions.
  
  The central part of ALICE covers $\pm~45^\circ$ ($|\eta| < 0.9$),
  and consists of an inner tracker (ITS), a TPC and a particle
  identification array (PID), all embedded in a large magnet with a
  weak solenoidal field.
  The experiment is completed by two small area spectrometers in the
  barrel region (an electromagnetic calorimeter, PHOS, and a high
  momentum PID detector, HMPID), a forward muon spectrometer
  (2$^\circ$ to 9.5$^\circ$) and a ZDC.

\end{minipage}
\end{center}

\end{titlepage}

\subsection{Introduction}

The aim of high-energy heavy-ion physics is the study of strongly
interacting matter at extreme energy densities.  Statistical QCD
predicts that, at sufficiently high energy densities, hadronic matter
melts into a plasma of deconfined quarks and gluons --- a transition
which took place, in the inverse direction, some $10^{-5}$~s after the
Big Bang and which might still play a role in the core of collapsing
neutron stars.  

Using methods and concepts from both nuclear and high-energy physics,
the study of the phase diagram of QCD matter and of quarks and gluons
in a deconfined medium, constitutes a new and interdisciplinary
approach in investigating the small scale structure of matter and its
interactions.  It explores and tests QCD on its natural scale
($\Lambda_\mathrm{QCD}$) and addresses the understanding of
confinement and chiral-symmetry breaking.

The current heavy ion program, in particular the new results obtained
with Pb beams at the SPS, gives clear indications that very dense
matter is produced in these reactions, the high energy densities
leading to new phenomena, beyond expectations from p--A physics.

At the LHC, besides the pp program, nucleus--nucleus collisions at a
centre-of-mass energy of about 6~TeV/nucleon are foreseen as a
prominent part of the initial experimental program.
Heavy-ion collisions at the LHC will certainly provide a suitable
environment for the study of strongly interacting matter.
Extrapolating from present results, the parameters relevant to the
formation of the Quark--Gluon Plasma (energy density, size and
lifetime of the system, etc.) will all be more favourable.
The {\it average} energy densities should be well above the
deconfinement threshold, and the QGP is expected to be probed in its
asymptotically free `ideal gas' form.  Unlike at lower energies, the
central rapidity region will have nearly vanishing baryon number
density, similar to the state of the early universe.

ALICE groups together people and experience from the community
presently engaged in the SPS heavy-ion program, joined by new groups
coming from both nuclear and high-energy physics.  It currently
includes around 600 physicists, from 65 institutions.
The ALICE technical proposal~\cite{tp} was submitted in December of
1995, providing a detailed description of the central detector.  As
was already the case for the Letter of Intent~\cite{loi, loimumu}, the
forward muon spectrometer will be described in a specific additional
document, currently under preparation and due by October of 1996.

ALICE is designed as a general-purpose heavy ion detector, sensitive
to the majority of known observables (including spectra of hadrons,
electrons, muons and photons) and will be operational at the LHC
start-up.  In addition to the heavier systems (Pb--Pb), ALICE will
study collisions of lower-mass ions (e.g. A$\,\sim40$), in order to
vary the energy density.  Reference data will be obtained from pp and
(when available) p--nucleus collisions.

\subsection{The ALICE detector}

\begin{figure}[htb]
\centering
\resizebox{0.9\textwidth}{!}{
\includegraphics*{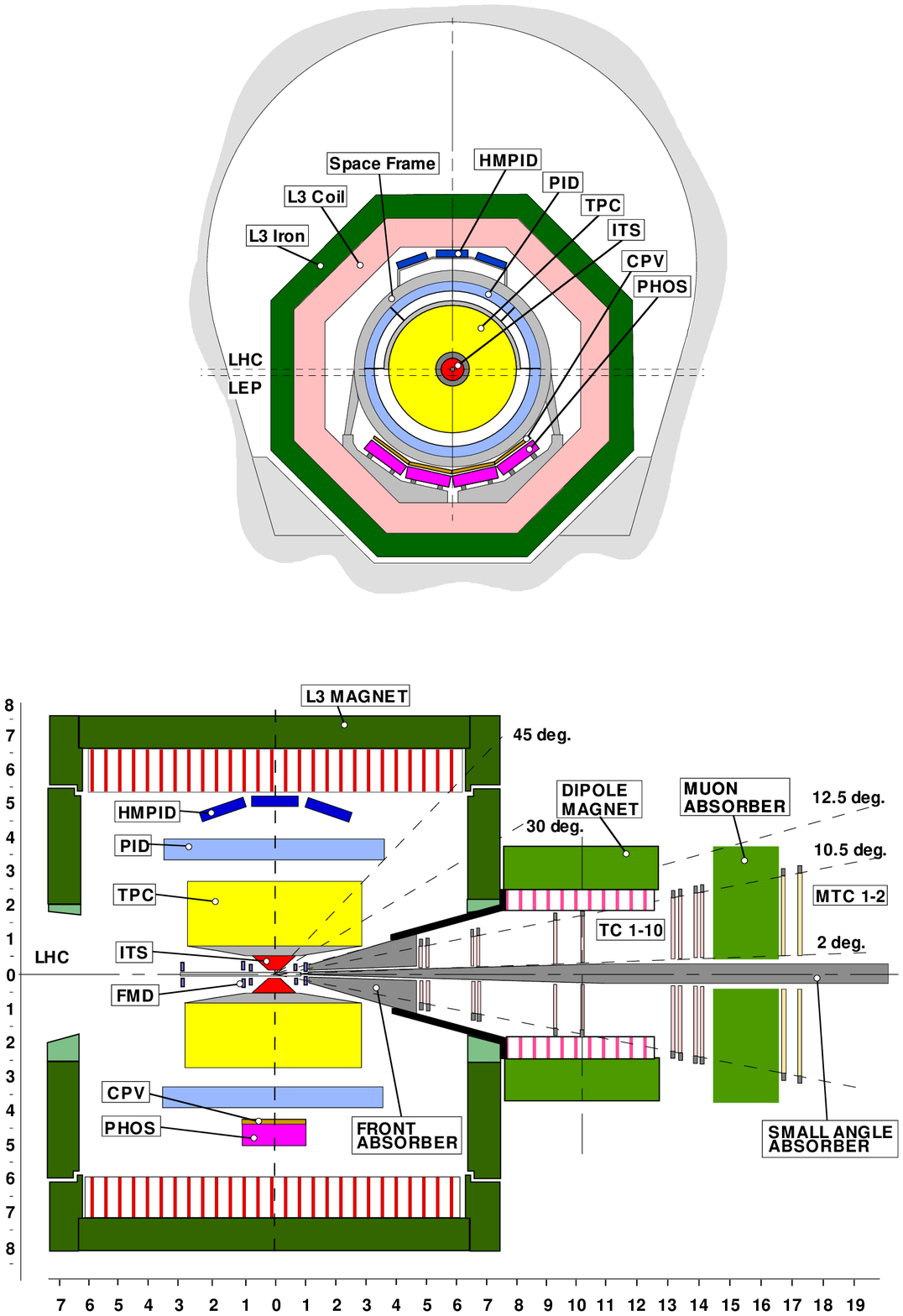}}
\caption{Longitudinal view of the ALICE detector.}
\label{fig:layout}
\end{figure}

The overall detector layout of the ALICE experiment is shown in
figure~\ref{fig:layout}.  The central part, covering $\pm~45^\circ$
($|\eta| < 0.9$) over the full azimuth, is embedded in a large magnet
with a weak solenoidal field (about 0.2~T) --- a compromise between
momentum resolution, low momentum acceptance and tracking efficiency
--- aiming at full tracking and particle identification.  It consists
of three cylindrical sub-detector layers: the inner tracker (ITS) with
six layers of high-resolution silicon tracking detectors, the TPC and
the particle identification array of TOF counters (PID).  Still in the
barrel region, there are two single-arm detectors: an electromagnetic
calorimeter (PHOS) and an array of counters optimized for
high-momentum inclusive particle identification (HMPID), made up
either of RICH or of TOF counters.  The forward muon spectrometer
covers the angular acceptance between 9.5$^\circ$ and 2$^\circ$ ($\eta
= 2.5 - 4.0$).  A zero-degree calorimeter, located some 92~m away from
the intersection point, will measure the energy of the spectator
nucleons and, therefore, access the collision geometry.  Finally, a
forward multiplicity detector (FMD) will extend the central rapidity
coverage into $|\eta|<4$.

The basic functions of the inner tracker ---\,secondary vertex
reconstruction of charm and hyperon decays, particle identification
and tracking of low-momentum particles, improvement of the momentum
resolution\,--- are achieved with six barrels of high-resolution
silicon detectors.  The number of layers and their radial position has
been optimized for efficient pattern recognition and impact parameter
resolution. Because of the high particle density, the four innermost
layers will be silicon pixel and silicon drift detectors. The two
outer layers will be equipped with silicon micro-strip detectors.
With the exception of the two innermost pixel planes, all layers will
have analog readout for independent particle identification via
d$E$/d$x$ in the non-relativistic region, allowing track finding of
low-\pt\ charged particles, down to a \pt\ of $\sim$\,20~MeV/$c$ for
electrons.

The need for efficient and robust tracking has led to the choice of a
TPC as the main tracking system.  In spite of its drawbacks concerning
speed and data volume, we have concluded that only a conventional
device and redundant tracking can guarantee reliable performance at up
to 8000 charged particles per unit of rapidity.  The inner radius of
the TPC ($r\approx90$~cm) is given by the maximum acceptable hit
density (0.1~cm$^{-2}$), while the outer radius (250~cm) is determined
by the length required for a d$E$/d$x$ resolution better than 7\,\%.
With such a resolution, the TPC can also be used to identify electrons
of momenta up to $\sim$\,2.5~GeV/$c$.  The track finding in the TPC
achieves an efficiency of close to 97\,\% for \pt\ as low as
100~MeV/$c$.

Particle identification over a large part of the phase space and for
many different particles is an important design feature of ALICE. It
is of crucial importance for a number of signals (\eg flavour
composition and chemical equilibrium, lepton pairs, strangelets) and
very useful for others (\eg momentum spectra, HBT, charm, jet
quenching). We have two detector systems dedicated exclusively to PID,
a TOF array optimized for large acceptance and average momenta and a
small system specialized on higher momenta.  Two TOF technologies,
Pestov spark counters and parallel plate chambers (PPC), are being
studied for the large area PID barrel, located at a radius of about
3.5~m.  A timing resolution of less than 50~ps (r.m.s.) has already
been achieved with first prototypes of Pestov counters.  For the PPCs,
corresponding values of $\lesssim$\,200~ps have been obtained.  A
second PID system, of smaller acceptance and at larger radii, will
extend the accessible momentum range for inclusive particle spectra
into the semi-hard region.  The current technological choice is a
proximity-focusing RICH detector with liquid freon radiator, solid
photocathode and pad readout.

Prompt photons, $\pi^{0}$'s and $\eta$'s will be measured in a
single-arm high-resolution electromagnetic calorimeter.  It is located
5~m from the vertex and will be built from PbWO$_4$ scintillating
crystals, a material with small Moli\`{e}re radius (2~cm), in order to
keep a reduced occupancy.  Since we are interested in accessing
relatively small energies ($\lesssim$\,15~GeV) the crystals will be
cooled to $-25^{\circ}$C, to increase the light output (the magnetic
field imposes silicon photodiodes readout).

One of the most promising signatures for the existence of deconfined
matter is the suppression of heavy quarkonia resonances.  The forward
muon arm will measure the dimuon decay of heavy quarkonia (\Jpsi,
\psip, \Ups, \Upsp\ and \Upspp) with a mass resolution sufficient to
separate all resonances.

During the heavy-ion run, two different types of events will be
collected.  In the first type, all detectors are read out, leading to
a huge volume of data (up to 40~MByte per event, at a rate of 50~Hz).
In the second type, specific for the physics of high-mass muon pairs,
only the muon arm and the pixel planes (for vertex finding) are read
out.  These triggers, interposed with the previous ones, have a much
higher rate (up to 1~kHz) but the event size is much smaller, the data
throughput being only 20\,\% of the total.  The very large data
volume, imposed by up to 12~000 charged tracks going through the TPC,
leads to $\sim$\,1000~TByte written during one month of heavy-ion
running (the pp LHC experiments reach a similar data volume after
\emph{one year}).

\subsection{Physics with the ALICE detector}

In order to establish and analyse the existence of QCD bulk matter and
the QGP, a number of observables have to be studied with ALICE in a
systematic and comprehensive way.  Some observables are needed to
characterize the global event features of the state created in the
nucleus--nucleus collision, to access the number of colliding nucleons
and the energy density reached.  This information is necessary to
interpret a specific signal as an indication of new physics.  We will
study several of these \emph{specific signals} together with
\emph{global information} about the events, in the same experiment.

The signals accessible to ALICE are described in detail in Chapter~11
of Ref.~\citen{tp}. They are summarized below according to the aspect
of the collision on which they have a bearing (see, for example,
Ref.~\citen{js} for original references).

%\subsubsection{Quark-Gluon Plasma}

The {\it thermal radiation} from both the QGP and the mixed phase will
be observable with the photon spectrometer in the medium-$p_{\rm t}$
range (around 1--3~GeV/$c$) if its rate is more than 5\,\% of the
(abundant) photons from hadronic decays.  At higher momenta, {\it
prompt photons} from the pre-equilibrium phase (3--6~GeV/$c$) contain
information on the parton dynamics at very early stages, on the
equilibration times and on the transition from perturbative to
non-perturbative phenomena.

The {\it light vector mesons}, with lifetimes of the order of the
expansion time scale, will partially decay during the evolution of the
system. Their properties (mass, width, branching ratios), observable
in the leptonic decay, should change in dense matter owing, for
example, to `collision broadening', `induced radiation' and other
in-medium effects. In the vicinity of the phase transition, partial
restoration of chiral symmetry will lead to additional shifts of the
mass. The excellent mass resolution of ALICE for electron pairs
($\Delta m/m \approx 1\%$) will allow to observe significant changes
for both the $\omega$ and $\phi$ mesons.

The suppression of {\it heavy quarkonia resonances} via Debye
screening is an important tool to diagnose deconfinement and the early
stages of the QGP. The muon spectrometer will measure the production
(and suppression) rates for the complete spectrum of heavy quarkonia.
\emph{Open charm} production will probe the parton kinematics in the
very early stage.  The cross-section of \emph{high}-\pt\
\emph{hadrons} is sensitive to the energy loss of the partons in the
plasma.

%\subsubsection{Phase transition}

\emph{Strangeness production} is sensitive to the large s-quark
density expected from (partial) chiral-symmetry restoration in the
plasma.  \emph{Multiplicity fluctuations} are a sign of critical
phenomena at the onset of a phase transition.  The expansion time in
the mixed phase, expected to be long in the case of a first-order
phase transition, can be measured by \emph{particle
interferometry}. High statistics data will be available at the LHC to
make a differential space-time analysis as a function of d$N$/dy and
$p_{\rm t}$.

%\subsubsection{Hadronic Matter}

A large number of {\it hadrons} ($\pi, \eta, \omega, \phi$, p, K,
$\Lambda$, $\Xi$, $\Omega$) will be measured as a function of
charged-particle density and $p_{\rm t}$. This will test {\it
thermalization} (equilibrium of particle ratios and momentum
distribution) and {\it dynamical evolution scenarios} in the hadronic
phase.  Enhanced production of {\it strangeness} from the QGP phase
might still be visible in the hadronic matter and can be searched for
with a few percent accuracy event-by-event.  The shape of the $p_{\rm
t}$ distribution and the average $p_{\rm t}$ will be measured for the
many pions, kaons and even protons on a event-by--event basis.
Therefore individual events can be assigned a `temperature' per
particle type which can be correlated with other observables.
Inclusive high-statistics measurements of hadrons will allow the
investigation of expansion dynamics and collective flow phenomena.

The freeze-out radius of the hadronic fireball can be measured by {\it
interferometry} on an event by event basis up to 15~fm, and
inclusively up to 30--40~fm.  The large radii and the long lifetimes
expected at LHC pose severe requirements on the detector in terms of
momentum and two-track resolution.

\bigskip

Among the most challenging measurements ALICE will perform are
di-electrons and open charm, both very demanding in terms of
statistics, in view of the small signal to background ratio.

The di-electron measurement requires very high efficiency to track and
identify electrons in order to recognize and eliminate the background
from Dalitz decays and conversions. The design of the ITS has been
optimized in this respect, in particular for low-\pt\ electrons (20 to
100~MeV/$c$), for which the outer detectors are less useful. 

\begin{figure}[t]
\centering
\begin{tabular}{cc}
\resizebox{0.4\textwidth}{0.25\textheight}{%
\includegraphics*{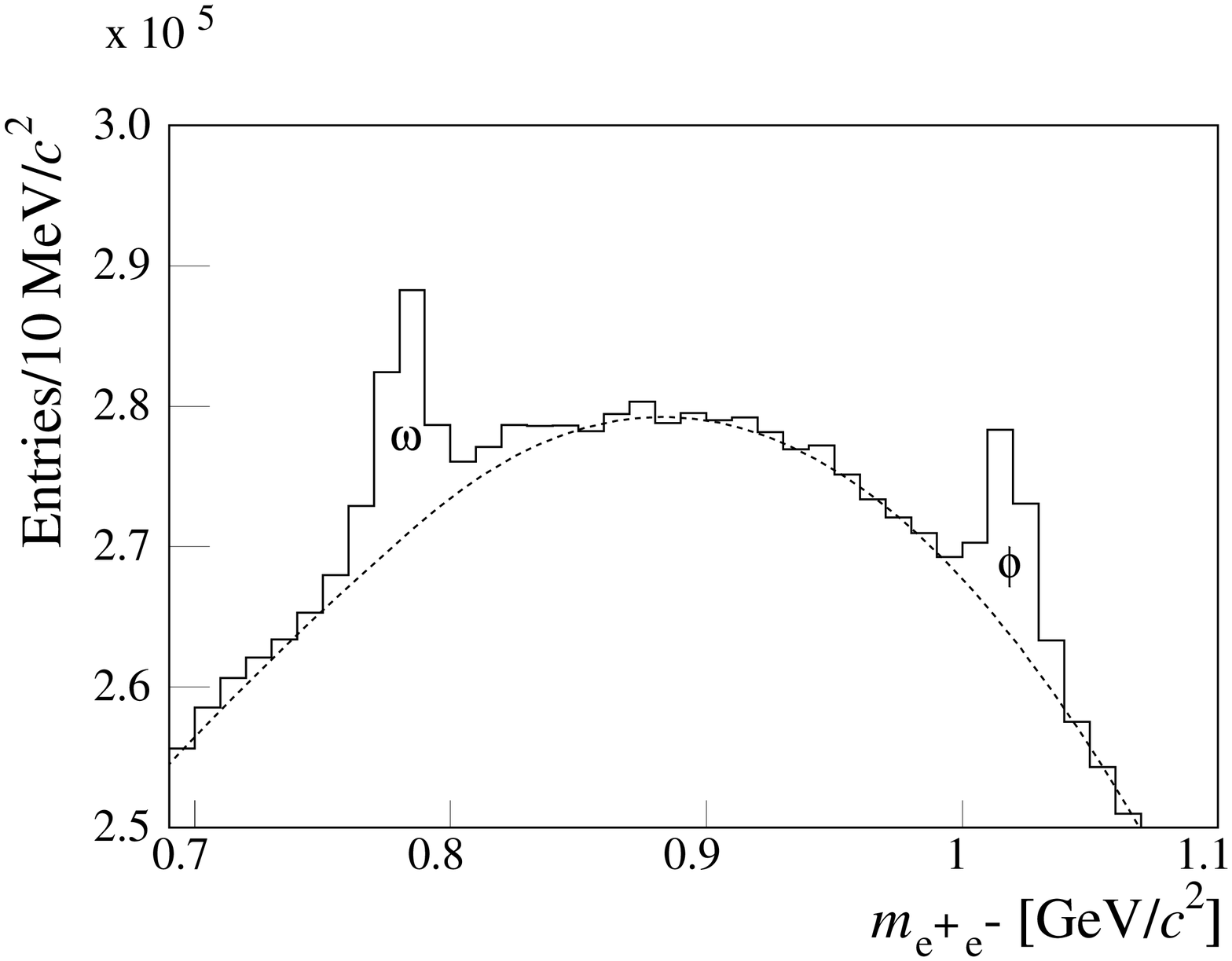}}
&
\resizebox{0.48\textwidth}{0.25\textheight}{%
\includegraphics*{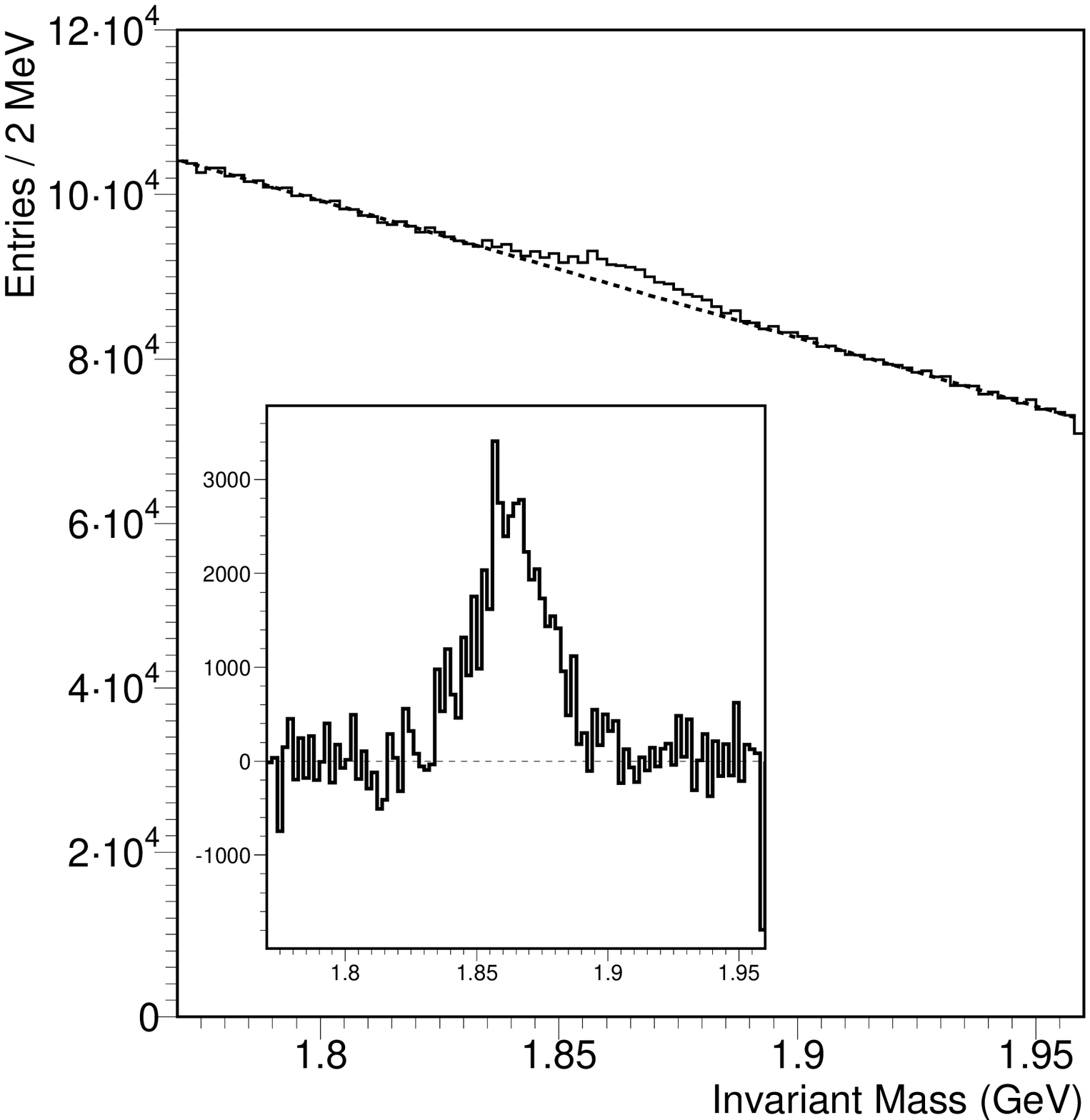}}
\end{tabular}
\caption{Left: Low mass di-electron spectrum, after cuts, for $5\times
  10^7$ central events, including tracking efficiency and momentum
  resolution.  Right: $\mathrm{K}^- \pi^+$ (+c.c.) invariant mass
  distribution before and after (inset) background subtraction.}
\label{fig:phys}
\end{figure}

The photon conversion background is greatly reduced by requiring a hit
in the first pixel layer, which is expected to have an efficiency
$\geq$\,98\,\%.  About 80\,\% of the conversions which still occur, in
the beam pipe (0.17\,\%~$X_0$) and in the first pixel layer itself
(0.15\,\%~$X_0$), can be recognized by reconstructing the secondary
vertex (conversion point).  The remaining conversions are well below
the level of electrons from Dalitz decays, which will also be reduced,
essentially by rejecting very low mass pairs
($m_{\rm{ee}}\lesssim100$~MeV).  Semileptonic charm decays are
suppressed by checking if the electron track points to the main
vertex.  The $\omega/\phi$ region of the expected invariant mass
distribution of pairs passing the selection cuts is shown in
figure~\ref{fig:phys}, where we can see the effect of having an
excelent mass resolution ($\sim$\,1\,\%).

Figure~\ref{fig:phys} also shows the invariant mass spectra of
$\mathrm{K}^- \pi^+$ (and $\mathrm{K}^+ \pi^-$), where the clearly
visible signal of D$^0$ production ($S/\sqrt{B} = 32$) corresponds to
a sample of $10^7$ events (about five days of running at 50\,\% data
taking efficiency), after quite stringent event selection cuts to
improve the signal to background ratio.  This measurement relies on a
very good impact parameter resolution, expected to be better than
100~$\mu$m, in the bending direction, for $\pi$ and K mesons of
\pt$\geq$\,600~MeV/$c$.  A very large gain in $S/B$ is also obtained
by requiring the momentum vector of the D$^0$ candidate to be aligned
with the line connecting the primary vertex to the candidate decay
vertex.

\subsection{Conclusion}

Heavy ion collisions at the LHC will open a new regime of very high
energy density but low baryon density.  With Pb--Pb collisions at a
$\sqrt{s}$ of more than 1200~TeV, we expect extreme particle densities
(several thousand per unit rapidity), systems approaching
100\,000~fm$^{3}$ and initial energy densities 50 to 100 times larger
than present in normal nuclear matter.  ALICE will be well prepared to
explore this `little Big Bang' and enter the wonderland of QCD
thermodynamics.

\end{document}